\documentclass[10pt, conference]{IEEEtran}
\IEEEoverridecommandlockouts

\usepackage{cite}
\usepackage{algorithmic}
\usepackage{graphicx}
\usepackage{textcomp}
\usepackage{xcolor}
\usepackage{amsmath,amssymb,amsfonts,amsthm}
\usepackage{physics}
\usepackage{dsfont}
\usepackage{comment}
\newcommand{\problem}[1]{$#1$-SYM}

\usepackage{bbm}

\usepackage{tikz}
\newcommand\copyrighttext{%
  \footnotesize \textcopyright 2025 IEEE. Personal use of this material is permitted.
  Permission from IEEE must be obtained for all other uses, in any current or future
  media, including reprinting/republishing this material for advertising or promotional
  purposes, creating new collective works, for resale or redistribution to servers or
  lists, or reuse of any copyrighted component of this work in other works.}
\newcommand\copyrightnotice{%
\begin{tikzpicture}[remember picture,overlay]
\node[anchor=south,yshift=10pt] at (current page.south) 
  {\fbox{\parbox{\dimexpr\textwidth-\fboxsep-\fboxrule\relax}{\copyrighttext}}};
\end{tikzpicture}%
}

\def\BibTeX{{\rm B\kern-.05em{\sc i\kern-.025em b}\kern-.08em
    T\kern-.1667em\lower.7ex\hbox{E}\kern-.125emX}}

\begin{document}

\title{Detecting Symmetrizability in Physical Systems}

\author{
    \IEEEauthorblockN{Florian Seitz, Janis N\"otzel \emph{(Member, IEEE)}}
    \thanks{This work was financed by the Federal Ministry of Education and Research of Germany via grants 16KISQ039 and 16KISQ077 as well as in the programme of ``Souver\"an. Digital. Vernetzt.''. Joint project 6G-life, project identification number: 16KISK002, by the Bavarian Ministry for Economic Affairs (StMWi) via the 6GQT project and by the DFG via grant NO 1129/2-1. Inspiring discussions with Prakash Narayan are acknowledged.}
    \IEEEauthorblockA{
        Emmy-Noether Gruppe Theoretisches Quantensystemdesign\\
        Technische Universit\"at M\"unchen\\
        \{flo.seitz, janis.noetzel\}@tum.de
    }
}

\maketitle
\copyrightnotice

\begin{abstract}
We study the problem of data transmission under the influence of a jammer, which is typical for wireless systems and commonly modeled as an arbitrarily varying channel (AVC) in information theory. AVC fulfilling a certain set of linear equations are called symmetrizable and are known to be prone to denial of service attacks.
Recent work has shown that deciding if a given AVC is symmetrizable or not is a non-Turing computable problem. 
By relaxing the formulation of symmetrizability, we show the existence of a polynomial-time algorithm that determines whether a given AVC is non-symmetrizable, but displays a critical dependence on the number of jammer input states. We then show how imposing an energy constraint on the jammer allows the same algorithm to efficiently identify large classes of AVCs which are non-symmetrizable.
\end{abstract}

\begin{IEEEkeywords}
Denial of Service Attack, Computability, Physical Layer
\end{IEEEkeywords}

\section{Introduction}
Arbitrarily Varying Channels (AVCs) model communication systems under jamming attacks. Unlike compound- and memoryless channels, they exhibit a rich behavior in the sense that a so-called \emph{symmetrizability condition} decides if the capacity is zero or equals the Shannon capacity of a specific channel in the convex hull of the set of channels defining the AVC \cite{ahlswede1978,Csiszar1988}. This condition takes the form of a set of equations
\begin{align}\label{def:symmetrizability}
\forall \, x,\hat{x},y : \; \sum_{s} W(y|s,x)U(s|\hat{x}) = \sum_{s}W(y|s,\hat{x})U(s|x)
\end{align}
wherein $W(y|s,x)$ is the probability that the receiver receives message $y$ if the sender sends $x$ and the jammer input is $s$. If a conditional probability distribution $U(s|x)$ can be found that satisfies \eqref{def:symmetrizability}, the AVC $W(y|s,x)$ is called \emph{symmetrizable}. The symmetrizability of AVCs has been explored in various previous works, such as \cite{Csiszar1988}, where Csisz\'{a}r and Narayan rigorously introduce the symmetrizability condition and its role in determining when the deterministic capacity of an arbitrarily varying channel is positive, following a foundational work by Ahlswede \cite{ahlswede1978}. The question of whether or not the respective linear-algebraic conditions are computable has been the subject of e.g. \cite{Boche19}, where the authors show that the question of whether or not an arbitrarily varying channel is symmetrizable is uncomputable. They also point out \cite[Theorem 3]{Boche19} the existence of a Turing machine which halts if a given AVC is non-symmetrizable. There exist two very distinct classes of AVCs - one is the class of discrete systems, where receiver, jammer and sender can only send and receive discrete symbols. The other is the class of continuous AVCs, which is studied for example in \cite{csiszarNarayan}, in the form of a Gaussian AVC with additive jamming. In this latter work it was proven that the capacity of the noiseless Gaussian AVC equals 
\begin{align}\label{eqn:capacity-for-Gaussian-AVC}
    C=\left\{\begin{array}{ll}\tfrac{1}{2}\log(1+E/P),&P<E\\0,&P\geq E\end{array}\right.
\end{align}
where $E$ and $P$ are the input power constraints of the sender and the jammer, respectively.

The striking difference between the two formulations \eqref{def:symmetrizability} and \eqref{eqn:capacity-for-Gaussian-AVC} as well as the observation of \cite{Boche19} that \eqref{def:symmetrizability} is in general not computable while \eqref{eqn:capacity-for-Gaussian-AVC} clearly is, and finally the fact that \eqref{def:symmetrizability} is a purely mathematical model, while \eqref{eqn:capacity-for-Gaussian-AVC} takes into account the underlying physics, motivates us to ask 

\emph{
How can the laws of physics guide the design of algorithms, such that the number of situations where we can decide within a finite time window if a system is safe to use, is increased?
}

We approach this question starting from the formulation \eqref{def:symmetrizability}, which we transform into a linear program that reveals within a predictable runtime whether the condition is \emph{approximately} fulfilled. If it is not approximately fulfilled, the communication system $W$ cannot be jammed. This natural formulation then reveals the devastating impact of the number $S$ of possible different inputs $s$ of the jammer on the time it takes to decide if the system can be jammed. Given that continuous systems such as the Gaussian AVC with additive jamming have infinitely many possible inputs for the jammer, this raises the question of how to design algorithmic decision procedures for such systems. Based on a standard modeling approach for an optical M-PSK system we investigate the symmetrizability of specific physical communication systems using numerical methods. For a more general case we then show how energy limitations such as the one in \eqref{eqn:capacity-for-Gaussian-AVC} can be utilized also for more complex communication systems as a tool to regain the ability of identifying systems which are ``safe to use'' (cannot be jammed).
\subsubsection*{Further Related Work} 
    In the early work \cite{relatedWork-narajanLapidoth} different AVC models are surveyed with emphasis on en- and decoders. The recent literature has introduced and studied the concept of myopic adversaries \cite{relatedWork-sarwate}, \cite{relatedWork-jaggi}, and studied the role of the AVC for receive diversity \cite{avcReceiveDiversity}. The Gaussian AVC was studied for broadcast systems in \cite{relatedWork-Kosut}. The relation to covert communication was explored in \cite{relatedWork-zhangCovert}. The detectability of DoS attacks was further studied in \cite{relatedWork-DoS}, and the impact of (non-) computability of certain functions was extended to a variety of domains in communications, including the computability of Fourier transforms \cite{relatedWork-ComputabilitySignalProcessing}. Arbitrarily varying quantum channels have been studied among others in \cite{relatedWork-quantumAVC,ahlswedeBlinovsky2007,reatedWork-ABBN}.

\section{Problem Statement}
Let $W$ be a classical channel depending on a parameter $s \in \mathcal S$ which is controlled by a jammer, so that $W(y|x,s)$ is the probability that the receiver gets the message $y \in \mathcal Y$ if the sender sent $x \in \mathcal X$ and the channel state was $s$. The input and output alphabets and the set of channel states are finite with $\abs{\mathcal X} = X, \abs{\mathcal Y} = Y$ and $\abs{\mathcal S} = S$. We call $\mathfrak{W} = \left\{ W(\cdot|\cdot,s) \right\}_{s \in \mathcal S}$ the arbitrarily varying channel. The strategy of the jammer is described by a channel $U$, where $U(s|\hat{x})$ is the probability that the jammer, who controls the channel state, uses $s$ if he picks a state $\hat{x}$. $\mathfrak W$ is called symmetrizable iff there exists a strategy $U$ such that
\begin{align}
\forall \, x,\hat{x},y : \; \sum_{s} W(y|x,s)U(s|\hat{x}) = \sum_{s} W(y|\hat{x},s)U(s|x).
\end{align}
In other words, the receiver can not decide if the sender sent $x$ and the jammer $\hat{x}$ or the other way around, which means communication is not possible \cite{ahlswede1978}.

For a given channel $\mathfrak W$ and set $\mathcal{X},\mathcal{Y},\mathcal{S}\subset \mathbb{N}$ the challenge is now to determine whether the function
\begin{align}\label{def:F(W)}
F(\mathfrak W) = &\min_{U} \max_{x \neq \hat{x}} \sum_{y \in \mathcal{Y} } \bigg| \sum_{s \in \mathcal{S} } W(y|x,s)U(s|\hat{x})  \\ &- \sum_{s \in \mathcal{S} }W(y|\hat{x},s)U(s|x) \bigg|,\nonumber \label{eq:F_function}
\end{align}
while also
\begin{align}
\forall x,s :&\sum_{y \in \mathcal{Y} } W(y|x,s) = 1 \\
\forall x: &\sum_{s \in \mathcal{S} }U(s|x) = 1,
\end{align}
is zero or has a positive value. In \cite{Boche19} it is shown that for all $X\geq2$, $S\geq2$, $Y\geq3$ there exists no Turing machine $\mathfrak T$ such that $\mathfrak T(\mathfrak W) = 1$ if and only if $F(\mathfrak W) = 0$.

\section{Approximate Symmetrizability}
Even though the exact problem is uncomputable, we can still make assertions about the symetrizability properties of an AVC. If we acknowledge for example that the detector on the receiver side inevitably works with finite precision, we may instead consider a channel to be $\varepsilon$-symmetrizable if $F(\mathfrak W) \leq \varepsilon$ for some $\varepsilon > 0$. Then the problem does become computable, which allows communicating parties to decide not to use a channel if $F(\mathfrak W)$ is too low. The computability can be seen by formulating the problem explicitly as a linear program.
\newpage
For that all we need to do is to introduce auxiliary variables $z(x, \hat x, y) \geq 0$ for every pair $(x,\hat x) \in \Tilde{\mathcal X}$ with $\smash{\Tilde{\mathcal X}} = \left\{ (x,\hat x) \in \mathcal X \times \mathcal X : x < \hat x \right\}$ and every $y \in \mathcal Y$ to linearize the absolute value by imposing the constraints
\begin{align}
    &z(x,\hat{x},y)  \\ &\geq \sum_{s\in\mathcal{S}} W(y|x,s) \, U(s|\hat{x}) - \sum_{s\in\mathcal{S}} W(y|\hat{x},s) \, U(s|x),\nonumber\\
    &z(x,\hat{x},y)  \\ &\geq -\left(\sum_{s\in\mathcal{S}} W(y|x,s) \, U(s|\hat{x}) - \sum_{s\in\mathcal{S}} W(y|\hat{x},s) \, U(s|x)\right).\nonumber
\end{align}
The remaining problem is already linear and we have
\begin{align}
    &\mathbf{Find}\nonumber\\ & \qquad \left\{U(s | x) \right\}_{s \in \mathcal S, \hat x \in \mathcal X}, \left\{ z(x, \hat x, y) \right\}_{(x, \hat x) \in \Tilde{\mathcal X}, y \in \mathcal Y}\\
    &\mathbf{Subject} \; \mathbf{to} \nonumber\\ & \qquad \forall x : \sum_{s \in \mathcal{S} }U(s|x) = 1\\
    & \qquad \forall s, x : U(s|x) \geq 0\\
    & \qquad \forall x < \hat x, y : z(x,\hat{x},y) \\ & \qquad \geq \sum_{s\in\mathcal{S}} W(y|x,s) \, U(s|\hat{x}) - \sum_{s\in\mathcal{S}} W(y|\hat{x},s) \, U(s|x)\nonumber\\
    & \qquad \forall x \neq \hat x, y : z(x,\hat{x},y) \\ & \qquad \geq -\left(\sum_{s\in\mathcal{S}} W(y|x,s) \, U(s|\hat{x}) - \sum_{s\in\mathcal{S}} W(y|\hat{x},s) \, U(s|x)\right)\nonumber\\
    & \qquad \forall x < \hat x: \sum_{y\in\mathcal{Y}} z(x,\hat{x},y) \leq \varepsilon
\end{align}
This linear feasibility problem is always computable and can even be solved in polynomial time. We call it the \problem{\varepsilon} problem.

\subsection*{Runtime Estimate}
The solvers employed for solving the \problem{\varepsilon} problem have runtime estimates in the order of
$\mathcal O((n+m)^\frac{3}{2} n L)$, where $n=\mathcal O(XS+X^2Y)$ is the number of variables, $m=\mathcal O (XS + X^2Y)$ is the number of constraints and  $L= \mathcal O(\log(\varepsilon( X^2Y + XS)))$ \cite{vaidya1989}, giving us a runtime on the order of $\mathcal O((XS+X^2Y)^\frac{5}{2}\log(\varepsilon(XS + X^2Y)))$.

Thus state of the art solvers will have problems to detect real-world DOS attacks where the jammer alphabet cannot be assumed to be finite. It thus turns out that even \problem{\varepsilon} may not be the right tool to analyze the impact of DOS attacks in systems beyond the scope of finite-alphabet information theory. In order to show that not all hope is lost, we show how to incorporate assumptions on the underlying physics to regain tractability.


\section{Symmetrizability of Random Channels}
\begin{figure}[h!]
    \centering
    \includegraphics[width=0.5\textwidth]{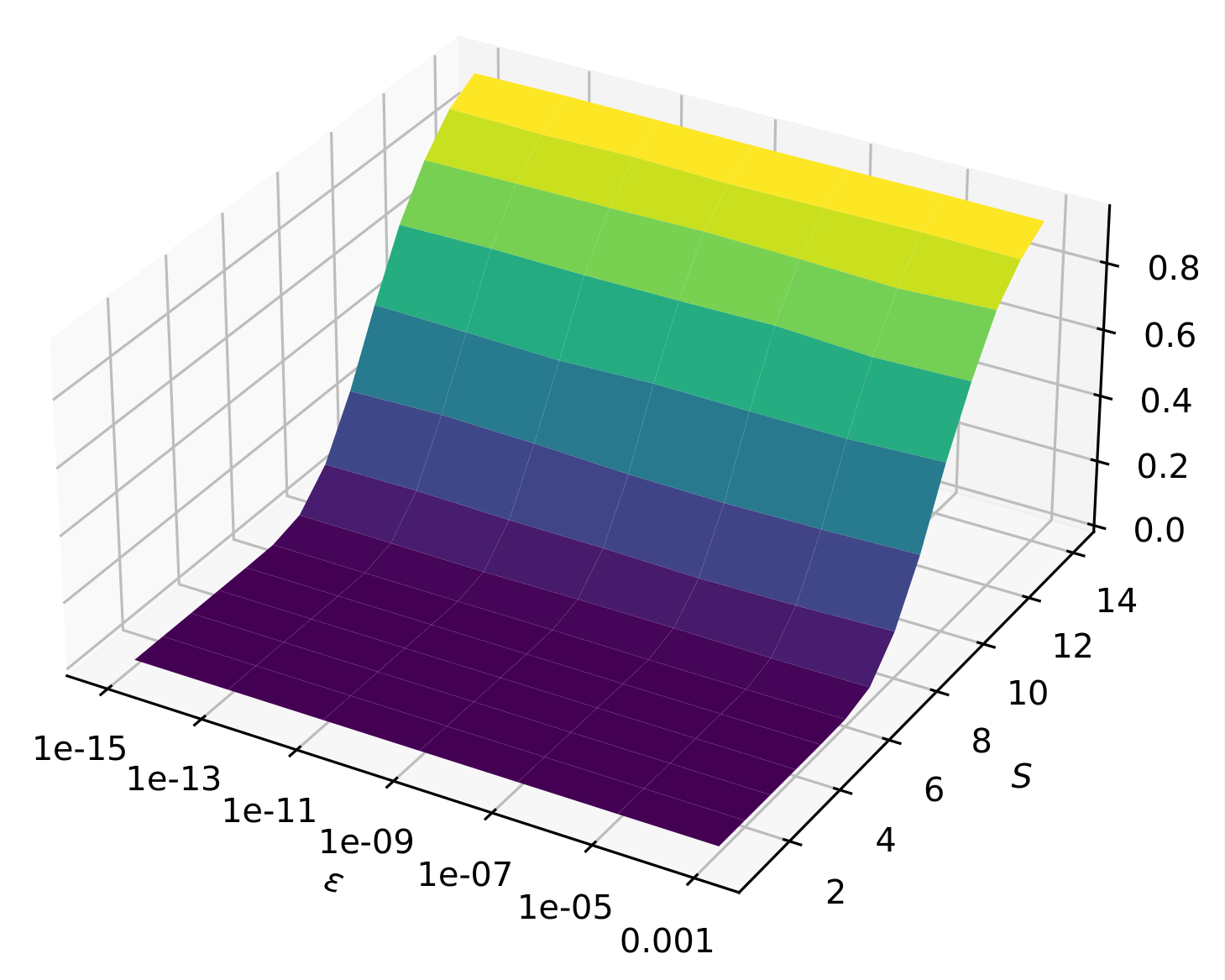}
    \vspace{-0.5cm}
    \caption{A surface plot showing the results of a numerical experiment to determine values of $p_\mathrm{sym}(X,Y,S,\varepsilon)$ for $X=Y=4$, $S = 2,...,14$, and $\varepsilon \in [2^{-15},2^{-3}]$. For each set of values, ten thousand samples were drawn.}
    \label{fig:surface}
\end{figure}

After realizing that the size of the jammer alphabet plays a crucial role in our ability to algorithmically determine if a specific AVC is symmetrizable, we would like to understand how likely it is that a random channel is $\varepsilon$-symmetrizable and how this is affected by the size of the jammer alphabet. For a fixed AVC $\mathfrak W$ of full rank the conditions
\begin{align}
    \sum_{s\in\mathcal{S}} W(y|x,s) \, U(s|\hat{x}) = \sum_{s\in\mathcal{S}} W(y|\hat{x},s) \, U(s|x).
\end{align}
form a system of $\frac{X(X-1)Y}{2}$ linear equations, while $U$ has $X(S-1)$ degrees of freedom. Given that the subset of rank-deficient matrices has Lebesgue measure zero within the space of all matrices of a given shape, we expect that almost no AVC is symmetrizable if $S-1 < \frac{(X-1)Y}{2}$. To test this, we perform a numerical experiment where we randomly generate an AVC $\mathfrak W$ and then determine $F(\mathfrak W)$. By drawing many samples for different values of $S$, we get an idea of how likely it is to find a symmetrizable channel up to a certain precision. Figure \ref{fig:surface} shows the results. The values of $X$ and $Y$ were fixed for all cases.
We start to see symmetrizable channels at $S=7$, and the ratio increases for larger values of $S$, just as expected. It is interesting to note that the proportion of $\varepsilon$-symmetrizable channels is independent of $\varepsilon$ within the parameter range of this experiment.

This analysis again highlights that if the size of the jammer alphabet is larger or even infinite, successful communication can not be expected in this generic case. Even though physical systems often admit continuous $\mathcal S$, they also pose constraints on the abilities of the jammer, enabling successful communication. In the remaining part of the paper, we investigate the symmetrizability properties of certain physical systems.

\section{A Physical Channel Model} \label{sec:physical_channel_model}
To develop a more concrete understanding of symmetrizability in a physical system, we analyze a specific channel model using tools from quantum optics. We derive the corresponding classical arbitrarily varying channels and numerically evaluate their symmetrizability. The model under consideration is a lossy bosonic channel with thermal noise, which is commonly used in fiber-optic communication. The interaction is modeled by a beam splitter: the sender and jammer control the two input ports, while the receiver observes one of the output ports. Signals are transmitted as displaced thermal Gaussian states, described by
\begin{align}
S_{\alpha}^{N} = \frac{1}{\pi N} \int_{\mathbb{C}} e^{-\frac{\left| \alpha-\mu \right| ^{2}}{N}}\ketbra{\mu}{\mu} \, d\mu = D(\alpha) S_{0}^{N} D^{\dagger}(\alpha),
\end{align}
and
\begin{align}
S_{0}^{N} = \frac{1}{N+1}\sum_{n=0}^{\infty} \left( \frac{N}{N+1} \right)^n \ketbra{n}{n},
\end{align}
where $\ket{n}$ denote Fock basis states, N is the number of thermal noise photons, and
\begin{align}
\ket{\mu} = e^{-\frac{\left| \mu \right| ^{2}}{2}	}\sum_{n=0}^{\infty} \frac{\mu^{n}}{\sqrt{ n! }} \ket{n}.
\end{align}
are coherent states.
Messages are encoded using an M-PSK scheme, meaning that a message $m \in \left\{ 1,\dots,M \right\}$ is encoded by a thermal state with displacement $\sqrt{ E } e^{2\pi i \frac{m}{M}}$, where $E$ is the power of the sender.
Sending two thermal states through a beam splitter creates a classically correlated state, but since we discard one of the outputs, the state at the receiver becomes again a simple thermal state. Let $N_{A}$ be the thermal noise of the sender and $N_{S}$ the thermal noise of the jammer. Then the output on the receiver side is
\begin{align}
\mathcal{N}(S_{\alpha}^{N_{A}}, S_{\beta}^{N_{S}}) = \mathrm{Tr}_{2}\left[ S_{\alpha}^{N_{A}} \boxplus_{\eta}  S_{\beta}^{N_{S}}  \right] = S_{\sqrt{ \eta }\alpha + \sqrt{ 1-\eta }\beta}^{\eta N_{A} + (1-\eta)N_{S}},
\end{align}
where $\boxplus_{\eta}$ denotes a beam splitter with transmittivity $\eta$. This can easily be seen from the Gaussian state representation of displaced thermal states and the beam splitter transformation (see for example \cite{brask2022}).
For the measurement, the receiver utilizes heterodyne detection \cite{rosati2017}. It is described by a continuous outcome POVM consisting of subnormalized coherent states,
\begin{align}
\left\{ \hat{E}_{\mu} = \frac{1}{\pi} \ketbra{\mu}{\mu } \right\} _{\mu \in \mathbb{C}},
\end{align}
which when applied to a thermal state $S_{\alpha}^N$ produces the outcome probability density
\begin{align}
p(x) = \mathrm{Tr}\left[\hat{E}_{x} \, S_{\alpha}^N \right] = \frac{1}{\pi (N+1)} e^{-\frac{\left| \alpha-x \right| ^{2}}{N+1}},
\end{align}
Finally, in the decoding step, the receiver must determine which message was sent based on the measurement outcome. To do this, the receiver compares the observed outcome to reference distributions corresponding to the expected states in the absence of jamming—i.e., when the jammer input is assumed to be a vacuum state with thermal noise. A maximum likelihood decoder selects the message whose reference state most likely produced the observed outcome.

\begin{figure}[h]
    \centering
    \includegraphics[width=0.5\textwidth]{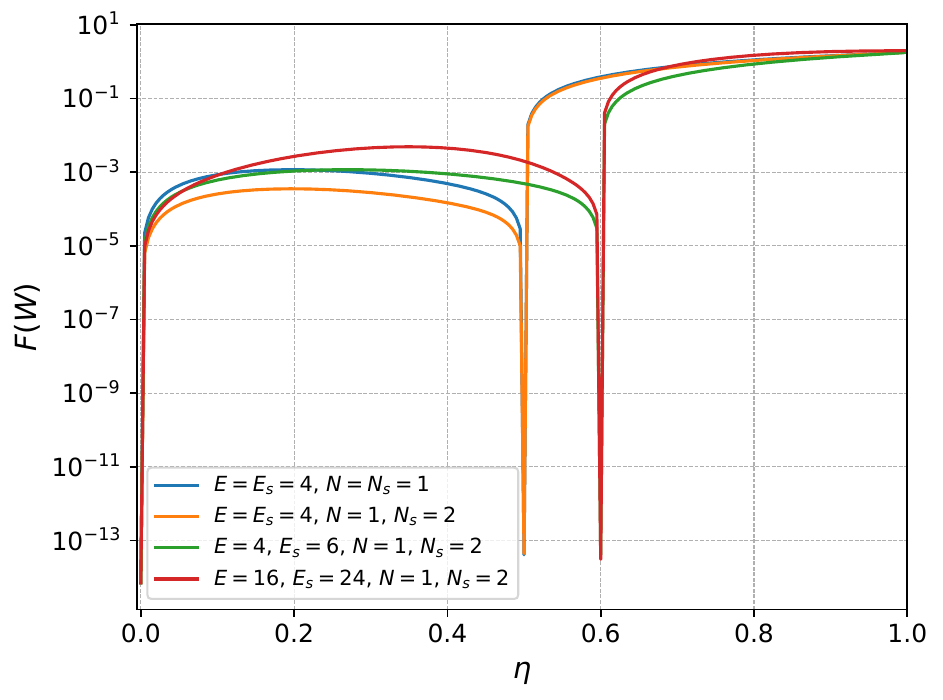}
    \vspace{-0.5cm}
    \caption{Minimal symmetrization error $F(\mathfrak W)$ according to \eqref{def:F(W)} of a lossy bosonic channel with thermal noise, for different transmittivities $\eta$. For all curves the number of messages is $M=6$.}
    \label{fig:physical_channel_sym.pdf}
\end{figure}

In the case of M-PSK encoding, this leads to a geometric interpretation: each message corresponds to a wedge-shaped acceptance region in the complex plane, bounded by the angles $\theta_{m}^{\pm} = 2\pi (m \pm \frac{1}{2})/M$, for the message $m$. These regions are fixed and do not depend on the level of thermal noise in the vacuum state. The probability of decoding message $m$ when the received state is $\rho$ is given by the integral of the measurement outcome distribution over the acceptance region corresponding to $m$,
\begin{align}
W(m|\rho) = \int_{\theta_{m}^-}^{\theta_{m}^{+}} d \theta \int_{0}^{\infty} dr \, r \, \mathrm{Tr}\left[\hat{E}_{r e^{i\theta}} \rho\right].
\end{align}
\begin{figure}[h]
    \centering
    \includegraphics[width=0.5\textwidth]{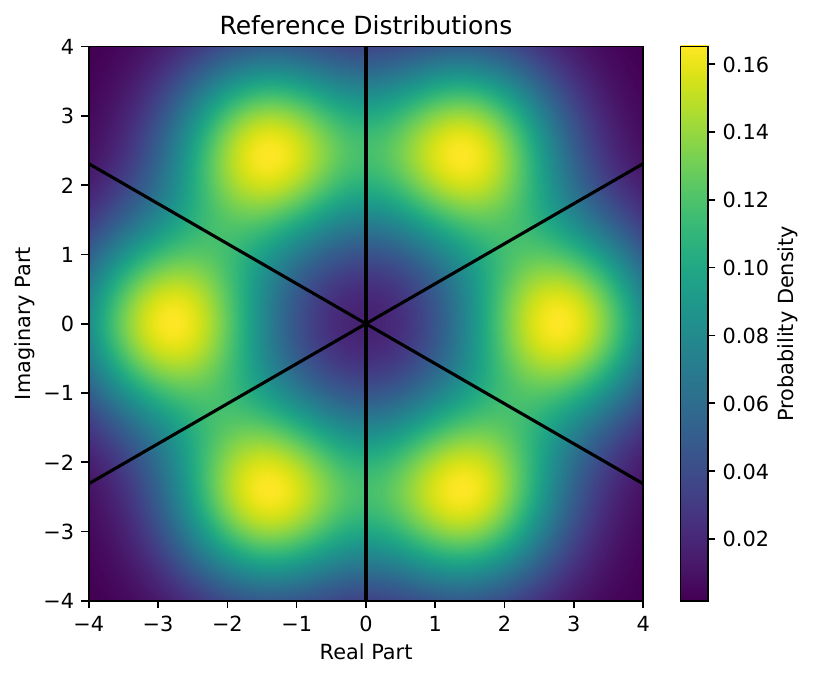}
    \vspace{-0.5cm}
    \caption{Visualization of the complex Gaussian outcome distributions for the different messages in case there is no jamming. The parameters in this case are $M=6$, $E=16$ and $N_A = N_S = 1$.}
    \label{fig:ref_dist}
\end{figure}
Figure \ref{fig:ref_dist} illustrates the structure of the reference distributions and the corresponding acceptance regions. The heatmap shows the sum of the measurement outcome distributions of the reference states, with black lines indicating the decision boundaries between messages.
In the simulation we assume that the jammer can transmit the same set of mesages as the sender, so $X = Y = S = M$. The channel is numerically evaluated for different values of $\eta$, the transmittivity of the beam splitter. For $\eta = \frac{1}{2}$, the effects of the sender and jammer states are exactly identical and we therefore expect the channel to be symmetrical by construction in this case. For each arbitrarily varying channel determined in this way, we numerically calculate the level to which it is symmetrizable, as defined in \eqref{eq:F_function}, by solving the linear optimization problem for the optimal jammer strategy. The results are plotted in Figure \ref{fig:physical_channel_sym.pdf}.
We observe that, regardless of the specific channel parameters, there are two points where exact symmetrization occurs, one for $\eta = 0$, in which case the inputs of the sender are just thrown out completely, and for a specific value that depends on the transmitter power and thermal noise of the sender and the jammer. This is the point at which both parties contribute the same amount of energy to the output signal, making the channel symmetrical by construction, as the heterodyne detector paired with a maximum likelihood decoder is unable to distinguish only based on thermal noise. In general, it can be stated that symmetrizability is the exception, and that even in situations where the jammer uses considerably more effective energy than the sender, this does not automatically lead to symmetrization of the channel.

Unlike in \cite{csiszarNarayan} (see \eqref{eqn:capacity-for-Gaussian-AVC}) where the jammer only needs to obey a \emph{maximum} power constraint, above model forces the jammer to use exactly the same strategy as the sender. It thus models a hardware-specific situation, a restriction which explains why the jammer's ability to carry out a DOS attack does not necessarily improve with the energy of the jammer's signals.

\section{Runtime Estimates under Energy Constraints\label{sec:runtime-estimates}}
Given the previous observations, it may appear that computational decision-making about the symmetrizability of real-world transmission systems where jammers have continuous degrees of freedom is out of reach as soon as one deviates from foundational models such as \cite{csiszarNarayan}. 
We therefore utilize our channel model from Section \ref{sec:physical_channel_model} to motivate a set of equations similar to \eqref{def:symmetrizability}, but with continuous degrees of freedom for the jammer, and then show how the assumption of limited jamming power lets us answer the \problem{\varepsilon} problem with a runtime that increases proportionally to the jamming power.
As the transition probabilities in our model, we take
\begin{align}
    w(y|s,x):=\Tr[M_y S_{\sqrt{ \eta }x + \sqrt{ 1-\eta }s}^{\eta N_{A} + (1-\eta)N_{S}}],
\end{align}
where $s,x\in\mathbb C$ and $\left\{M_y \right \}$ is a POVM. 
For such system with continuous degrees of freedom for the jammer and hard power limits per transmission, 
the symmetrizability condition \eqref{def:symmetrizability} can be rewritten as 
\begin{align}\label{eqn:continuous-symmetrizability-condition}
    \forall x,\hat x,y: \int [u(s|x)w(y|s,\hat x)- u(s|\hat x)w(y|s,x)]ds=0,
\end{align}
where $u(\cdot|x)$, $x\in X$, are probability density functions and $w(y|s,x)$ are continuous functions. 
The implication that the continuous symmetrizability condition \eqref{eqn:continuous-symmetrizability-condition} implies zero capacity follows trivially by following the lines of proof in \cite{Csiszar1988,ericson1985,ahlswedeBlinovsky2007} where in particular \cite[proof of Statement 1]{ahlswedeBlinovsky2007} provides a quick introduction to the relevant technique. The reverse statement, that non-symmetrizability according to \eqref{eqn:continuous-symmetrizability-condition} implies positive capacity, is less obvious and proven by us for a specific jamming attack in the appendix.
The problem can then again be relaxed by considering the function
\begin{align}
    f_{u,\mathfrak W}^{x,\hat x, y}&:=\|\int [u(s|x)w(y|s,\hat x)- u(s|\hat x)w(y|s,x)]ds\|,\label{eqn:fxxu}\\
    \mathcal F(\mathfrak W)&:=\max_{x,\hat x, y}\min_u f_{u,\mathfrak W}^{x,\hat x, y},
\end{align}
and requiring an algorithm which checks, for a given $\varepsilon>0$, whether $\mathcal F(\mathfrak W)\leq\varepsilon$. Relating the physical system to algorithmic implementation, we assume again the problem of an energy-constrained jammer whose input symbols $\beta$ satisfy $|\beta|^2\leq E_S$. For such a system, it must hold $u(s|x)=0$ whenever $|s|^2> E_S$ in \eqref{eqn:fxxu}. Let now
each jammer input be approximated to within distance $\delta$ by using a number $S_\delta \approx \frac{4 E_S}{\delta^2}$ of discrete points $\beta_1,\beta_2,\ldots,\beta_S$ which all satisfy $|\beta_i|^2\leq E_S$ and are arranged in a regular grid, so that the half-open rectangular boxes $\Box_i(\delta)$ around $\beta_i$ satisfy $\Box_i(\delta)\cap\Box_j(\delta)$ for $j\neq i$. 
For such a finite set of jammer states we define a discretization of the channel and the jammer strategy
\begin{align}
    \bar u(i|x):=&\int_{\Box_i(\delta)}u(s|x)ds \label{eq:discrete_u},\\
    \bar w(y|i,x):=& \frac{1}{\mu(\Box_i(\delta))} \int_{\Box_i(\delta)}w(y|s,x)ds.\label{eq:averaged_w}
\end{align}
We call the resulting discrete AVC $\overline{\mathfrak W}_\delta = \left\{\bar{w}(\cdot|i,\cdot)\right\}_{i=1}^{S_\delta}$ and we would like to show that the continuous AVC is approximately symmetrizable if and only if the same is true for the discrete AVC, in other words for every $\eta > 0$ there is a $ \delta > 0$ such that$ \abs{\mathcal F(\mathfrak W) - F(\overline{\mathfrak W}_\delta)} \leq \eta$.
We define
\begin{align}
    f_{\bar{u},\overline{\mathfrak W}_\delta}^{x,\hat x,y}:=\left|\sum_{i } [\bar{u}(i|x)\bar{w}(y|i,\hat x)- \bar{u}(i|\hat x)\bar{w}(y|i,x)]\right|,
\end{align}
and $F(\overline{\mathfrak W}_\delta)=\max_{x,\hat x,y}\min_{\bar{u}} f_{\bar{u},\overline{\mathfrak W}_\delta}^{x,\hat x,y}$, then the condition $\forall x,\hat x,y: \abs{f_{u,\mathfrak W}^{x,\hat x, y}-f_{\bar{u},\overline{\mathfrak W}_\delta}^{x,\hat x,y}} \leq \eta$ implies $\abs{\mathcal F(\mathfrak W) - F(\overline{\mathfrak W}_\delta)} \leq \eta$.
We have
\begin{align}
 & \left| f_{u,\mathfrak W}^{x,\hat x, y}-f_{\bar{u},\overline{\mathfrak W}_\delta}^{x,\hat x,y}  \right| \\
\leq  & \Big| \int [u(s|x)w(y|s,\hat x)ds -  \sum_{i} [\bar{u}(i|x)\bar{w}(y|i,\hat x) \nonumber \\
  -  &  \int [u(s|\hat{x})w(y|s,x)ds -  \sum_{i} [\bar{u}(i|\hat{x})\bar{w}(y|i, x) \Big| . \nonumber
\end{align}
Since the two terms are equivalent we would like to bound
\begin{align}
 & \left| \int [u(s|x)w(y|s,\hat x)ds -  \sum_{i} [\bar{u}(i|x)\bar{w}(y|i,\hat x) \right|  \\
\leq  & \sum_{i} \left| \int_{\Box_{i}(\delta)}u(s|x)w(y|s,x)ds - \bar u(i|x)\bar w(y|i,x) \right|.\nonumber
\end{align}
The function $w$ is continuous and defined on a compact set, therefore uniformly continuous, which implies that for every $\eta'>0$ there exists a $\delta$ such that
\begin{align}
\left| s-s' \right|\leq \delta \implies  \left| w(y|x,s)-w(y|x,s') \right| \leq \eta,
\end{align}
and as a consequence
\begin{align}
    s \in \Box_{i}(\delta) \implies \left| w(y|x,s) - \bar{w}(y|i,x) \right| \leq \eta'.
\end{align}
Then
\begin{align}
\textstyle\sum_{i}& \big| \int_{\Box_{i}(\delta)}u(s|x)w(y|s,x)ds - \bar u(i|x)\bar w(y|i,x) \big| \\
\textstyle \leq &  \sum_{i} \big| \eta' \int_{\Box_{i}(\delta)}u(s|x)ds \\ & + \int_{\Box_{i}(\delta)}u(s|x)\bar w(y|i,x)ds - \bar u(i|x)\bar w(y|i,x) \big| \nonumber \\
\textstyle \leq  & \eta' \sum_{i} \big| \int_{\Box_{i}(\delta)}u(s|x)ds\big| \\
= & \eta',
\end{align}
and $\left| f_{u,\mathfrak W}^{x,\hat x, y}-f_{\bar{u},\overline{\mathfrak W}_\delta}^{x,\hat x,y}  \right| \leq 2 \eta'$.

In the channel model we are considering here, the continuity of $w$ is due to the continuity of the channel and the linearity of the trace. We want to highlight, however, that the calculation works for any system for which $w$ is uniformly continuous in the jammer state, which covers a wide range of physical systems.
This analysis shows that finite approximations can be used as an efficient tool for analyzing the stability of nontrivial communication systems under DOS attacks, if these attacks can be bounded in energy.

\section{Conclusion}
We have defined \problem{\varepsilon} as a method of identifying, in deterministic time, whether a set of transition probabilities $w(y|s,x)$ describes a system with side channels which cannot be corrupted by a DOS attack. We highlighted the importance of tracking system parameters, in particular those describing the jammer's abilities, and defined a communication model with a jammer whose possible number of input states can be assumed as infinite, and defined a new \emph{continuous} symmetrizability condition for this model. We showed that we can efficiently compute this condition under the assumption of finite jamming power.

\section{Appendix}
Assume \eqref{eqn:continuous-symmetrizability-condition} is not true. Then there are $x_0,x_1, y$ such that 
\begin{align}\label{eqn:continuous-symmetrizability-condition-not-fulfilled}
    |\int [u(s|x_0)w(y|s,x_1)- u(s|x_1)w(y|s,x_0)]ds|>\epsilon
\end{align}
holds for all conditional probability densities $u(\cdot|x_i)$. Consider a code consisting of two messages $m_0,m_1$ with respective permutation-invariant length-$k$ code-words $x_0^k=(x_0,\ldots,x_0)$ and $x_1^k(x_1,\ldots,x_1)$. According to \eqref{eqn:continuous-symmetrizability-condition-not-fulfilled}, the convex sets 
\begin{align}
    A_i:=\mathrm{conv}(\{w(\cdot|s,x_i)\}_{|s|^2\leq E})
\end{align}
are disjoint. For every $k\in\mathbb N$ and $\xi>0$, we define $T_i$ to be the set of $y^k$ with the property that there exists $q\in A_i$ satisfying $\|\tfrac{1}{k} N(y|y^k) - q\|_1\leq\xi$.
For small enough $\xi>0$ and large enough $k\in\mathbb N$ Pinsker's inequality guarantees for $i=0,1$
\begin{align}
    T_i\cap T_{i\oplus1} = \emptyset,\qquad q^{\otimes k}(T_i)<2^{-k\xi^2/4} \;\forall q\in A_{i\oplus1}.
\end{align}
We take $T_i$ as decoding set for $x_i^k$. Since the resulting code is permutation-invariant, any jamming strategy $\rho^k=\rho_1\otimes\ldots\otimes \rho_k$ is equivalent to a corresponding strategy 
\begin{align}
    \bar\rho = \tfrac{1}{k!}\textstyle\sum_{\pi\in S_k}U_\pi \rho^k U_\pi^\dag,
\end{align}
where $S_k$ is the group of permutations of $k$ symbols and $U_\pi$ the representation of $\pi$. Since each $\rho_i$ can be written as 
\begin{align}
    \rho_i = \textstyle\int_{|\alpha|^2\leq E}p(\alpha)|\alpha\rangle\langle\alpha| d\alpha, 
\end{align}
we can take a finite subset $\beta_1,\ldots,\beta_S$ satisfying $|\beta_i|^2\leq E_S$ and $\alpha\in\cup_i\Box_i(\delta)$ whenever $|\alpha|^2\leq E_S$, as in Section \ref{sec:runtime-estimates}. We then know that with $\tilde p(i):=p(\Box_i(\delta))$ we get
\begin{align}
    \|\rho_i& - \textstyle\sum_{i=1}^S\tilde p(i)|\beta_i\rangle\langle\beta_i|\|_1
        \leq 2(1-e^{-2\delta^2}),
\end{align}
due to the equality $\||\alpha\rangle\langle\alpha|-|\beta\rangle\langle\beta|\|_1=1-e^{-|\alpha-\beta|^2}$.
Let the jamming sequence $\alpha^k$ satisfy $\max_{1\leq i\leq k}|\alpha_i|^2\leq E_S$. Define
\begin{align}
    \rho^k := \tfrac{1}{k!}\textstyle\sum_{\pi\in S_k}U_\pi\left(\otimes_{i=1}^k|\alpha_i\rangle\langle\alpha_i|\right) U_\pi^\dag.
\end{align}
To every $\alpha_i$, pick a corresponding $\tilde\alpha_i\in\{\beta_1,\ldots,\beta_S\}$ and set 
\begin{align}
    \tilde\rho^k := \tfrac{1}{k!}\textstyle\sum_{\pi\in S_k}U_\pi\left(\otimes_{i=1}^k|\tilde\alpha_i\rangle\langle\tilde\alpha_i|\right) U_\pi\dag.
\end{align}
Then 
\begin{align}
    \|\rho^k-\tilde\rho^k\|\leq k\cdot 2(1-e^{-2\delta^2}).
\end{align}
We now rewrite $\tilde\rho^k$, by letting $N(i|\tilde\alpha^k)$ be the number of times the symbol $\beta_i$ occurs in $\tilde\alpha^k$ and $p(i):=\tfrac{1}{k}N(i|\tilde\alpha^k)$. Then
\begin{align}
    \tilde\rho^k 
        = \frac{1}{p^{\otimes k}(T_N)}\textstyle\sum_{\beta^k\in T_N}p^{\otimes k}( \beta^k)\otimes_{i=1}^k|\beta_i\rangle\langle\beta_i|,
\end{align}
where $\tilde\beta^k$ is any element taken from $T_N$. It follows that 
\begin{align}
    \tilde\rho^k 
        & \leq (2k)^S\tilde\rho^{\otimes k},
\end{align}
where $\tilde\rho:=\sum_{i}p(\beta_i)|\beta_i\rangle\langle\beta_i|$. Note that $\tilde\rho$ has positive P representation and obeys the power constraint. We get 
\begin{align}
    w^{\otimes k}&(y^k|\alpha^k,x_j)=\Tr(\left(\otimes_{i=1}^kM_{y_i}S_{\sqrt{ \eta }x_j + \sqrt{ 1-\eta }s_i}^{\eta N_{A} + (1-\eta)N_{S}}\right))\\
        &\leq \Tr(\otimes_iM_{y_i}S_{\sqrt{ \eta }x_j + \sqrt{ 1-\eta }\tilde\alpha_i}^{\eta N_{A} + (1-\eta)N_{S}}) + k\cdot 2(1-e^{-2\delta^2})\\
        &= (2k)^Sq^{\otimes k}(y^k) + k\cdot 2(1-e^{-2\delta^2})
\end{align}
for some $q\in A_j$. The same bound applies if we replace $y^k$ by the sum over all $y^k$ in $T_{i\oplus1}$:
\begin{align}
    w^{\otimes k}(T_{j\oplus 1}|\alpha^k,x_j)\leq (2k)^Sq^{\otimes k}(T_{j\oplus 1}) + k\cdot 2(1-e^{-2\delta^2}).\nonumber
\end{align}
It hence suffices to pick $\delta=\delta_k$ such that $\lim_{k\to\infty}k\cdot 2(1-e^{-2\delta^2})=0$ and $\lim_{k\to\infty}(2k)^Sq^{\otimes k}(T_{j\oplus 1})=0$, and choosing $\delta_k=k^{-2/3}$ yields the desired behavior. Therefore a finite $k\in\mathbb N$ exists for which $T_0,T_1$ yield decoding error $<1/4$. Since we assume a maximum power constraint on the jammer, established techniques \cite{ahlswede1978} this show that the capacity of the AVC $w$ is nonzero.

\bibliographystyle{IEEEtran}
\bibliography{bib}

@INPROCEEDINGS{Boche19,
  author={Boche, Holger and Schaefer, Rafael F and Vincent Poor, H.},
  booktitle={ICASSP 2019 - 2019 IEEE International Conference on Acoustics, Speech and Signal Processing (ICASSP)}, 
  title={Detectability of Denial-of-service Attacks on Communication Systems}, 
  year={2019},
  volume={},
  number={},
  pages={2532-2536},
  doi={10.1109/ICASSP.2019.8683553}}

@misc{rosati2017,
      title={Decoding Protocols for Classical Communication on Quantum Channels}, 
      author={Matteo Rosati},
      year={2017},
      eprint={1710.08638},
      archivePrefix={arXiv},
      primaryClass={quant-ph},
      url={https://arxiv.org/abs/1710.08638}, 
}

@ARTICLE{Csiszar1988,
  author={Csisz\'{a}r, I. and Narayan, P.},
  journal={IEEE Transactions on Information Theory}, 
  title={The capacity of the arbitrarily varying channel revisited: positivity, constraints}, 
  year={1988},
  volume={34},
  number={2},
  pages={181-193},
  doi={10.1109/18.2627}}

@misc{brask2022,
      title={Gaussian states and operations -- a quick reference}, 
      author={Jonatan Bohr Brask},
      year={2022},
      eprint={2102.05748},
      archivePrefix={arXiv},
      primaryClass={quant-ph},
      url={https://arxiv.org/abs/2102.05748}, 
}

@article{ahlswede1978,
    author ={Ahlswede, Rudolf} ,
    title = {Elimination of correlation in random codes for arbitrarily varying channels},
    journal = {Zeitschrift für Wahrscheinlichkeitstheorie und Verwandte Gebiete},
    year = {1978},
    volume = {44},
    pages = {159-175}
}

@inproceedings {vaidya1989,
  author={Vaidya, P.M.},
  booktitle={30th Annual Symposium on Foundations of Computer Science}, 
  title={Speeding-up linear programming using fast matrix multiplication}, 
  year={1989},
  volume={},
  number={},
  pages={332-337},
  doi={10.1109/SFCS.1989.63499}}

@ARTICLE{ericson1985,
  author={Ericson, T.},
  journal={IEEE Transactions on Information Theory}, 
  title={Exponential error bounds for random codes in the arbitrarily varying channel}, 
  year={1985},
  volume={31},
  number={1},
  pages={42-48},
  doi={10.1109/TIT.1985.1056995}}

@ARTICLE{ahlswedeBlinovsky2007,
  author={Ahlswede, Rudolf and Blinovsky, Vladimir},
  journal={IEEE Transactions on Information Theory}, 
  title={Classical Capacity of Classical-Quantum Arbitrarily Varying Channels}, 
  year={2007},
  volume={53},
  number={2},
  pages={526-533},
  doi={10.1109/TIT.2006.889004}}

@ARTICLE{csiszarNarayan,  
    author={Csiszar, I. and Narayan, P.},  
    journal={IEEE Transactions on Information Theory},   
    title={Capacity of the Gaussian arbitrarily varying channel},   
    year={1991},  
    volume={37},  
    number={1},  
    pages={18-26},  
    doi={10.1109/18.61125}
}

@ARTICLE{relatedWork-jaggi,
  author={Zhang, Yihan and Vatedka, Shashank and Jaggi, Sidharth and Sarwate, Anand D.},
  journal={IEEE Transactions on Information Theory}, 
  title={Quadratically Constrained Myopic Adversarial Channels}, 
  year={2022},
  volume={68},
  number={8},
  pages={4901-4948},
  doi={10.1109/TIT.2022.3167554}}

@INPROCEEDINGS{relatedWork-sarwate,
  author={Sarwate, Anand D.},
  booktitle={2010 IEEE Information Theory Workshop}, 
  title={Coding against myopic adversaries}, 
  year={2010},
  volume={},
  number={},
  pages={1-5},
  doi={10.1109/CIG.2010.5592896}}

@ARTICLE{relatedWork-zhangCovert,
  author={Zhang, Qiaosheng and Bakshi, Mayank and Jaggi, Sidharth},
  journal={IEEE Transactions on Information Theory}, 
  title={Covert Communication Over Adversarially Jammed Channels}, 
  year={2021},
  volume={67},
  number={9},
  pages={6096-6121},
  doi={10.1109/TIT.2021.3096176}}

@INPROCEEDINGS{relatedWork-Kosut,
  author={Hosseinigoki, Fatemeh and Kosut, Oliver},
  booktitle={2020 IEEE International Symposium on Information Theory (ISIT)}, 
  title={Capacity Region of the Gaussian Arbitrarily-Varying Broadcast Channel}, 
  year={2020},
  volume={},
  number={},
  pages={1007-1011},
  doi={10.1109/ISIT44484.2020.9174108}}

@ARTICLE{relatedWork-ComputabilitySignalProcessing,
  author={Boche, Holger and Mönich, Ullrich J.},
  journal={IEEE Transactions on Signal Processing}, 
  title={Turing Computability of Fourier Transforms of Bandlimited and Discrete Signals}, 
  year={2020},
  volume={68},
  number={},
  pages={532-547},
  doi={10.1109/TSP.2020.2964204}}

@ARTICLE{relatedWork-DoS,
  author={Boche, Holger and Schaefer, Rafael F. and Poor, H. Vincent},
  journal={IEEE Transactions on Signal Processing}, 
  title={Denial-of-Service Attacks on Communication Systems: Detectability and Jammer Knowledge}, 
  year={2020},
  volume={68},
  number={},
  pages={3754-3768},
  doi={10.1109/TSP.2020.2993165}}

@article{relatedWork-narajanLapidoth,
  title={Reliable Communication Under Channel Uncertainty},
  author={Amos Lapidoth and Prakash Narayan},
  journal={IEEE Trans. Inf. Theory},
  year={1998},
  volume={44},
  pages={2148-2177},
  url={https://api.semanticscholar.org/CorpusID:14382189}
}

@INPROCEEDINGS{relatedWork-quantumAVC,
  author={Boche, Holger and Deppe, Christian and Nötzel, Janis and Winter, Andreas},
  booktitle={2018 IEEE International Symposium on Information Theory (ISIT)}, 
  title={Fully Quantum Arbitrarily Varying Channels: Random Coding Capacity and Capacity Dichotomy}, 
  year={2018},
  volume={},
  number={},
  pages={2012-2016},
  doi={10.1109/ISIT.2018.8437610}}

@article{reatedWork-ABBN,
    author= {Ahlswede, Rudolf and Bjelakovi\'c, Igor and Boche, Holger and Nötzel, Janis},
    year = {2013},
    title = {Quantum Capacity under Adversarial Quantum Noise: Arbitrarily Varying Quantum Channels},
    journal = {Communications in Mathematical Physics},
    pages = {103--156},
    volume = {317},
    issue = {1},
    doi = {10.1007/s00220-012-1613-x}
}

@article{avcReceiveDiversity,
    author = {Arendt, C. and Nötzel, J. and Boche, H.},
    title = {Reliable Communication under the Influence of a State-Constrained Jammer: An Information-Theoretic Perspective on Receive Diversity},
    journal = {Problems of Information Transmission},
    volume = {55}, 
    pages = {101--123},
    year = {2019},
    doi = {10.1134/S0032946019020017}
}
\end{document}